\documentclass[english,nofootinbib,notitlepage,superscriptaddress]{revtex4-1}
\usepackage[T1]{fontenc}
\usepackage[utf8]{inputenc}
\usepackage{geometry}
\usepackage{color}
\usepackage{babel}
\usepackage{amsmath}
\usepackage{amssymb}
\usepackage{graphicx}
\usepackage{esint}
\usepackage{comment}
\usepackage[unicode=true,pdfusetitle,
 bookmarks=true,bookmarksnumbered=false,bookmarksopen=false,
 breaklinks=false,pdfborder={0 0 0},backref=false,colorlinks=true,citecolor=blue]
 {hyperref}

\makeatletter
\usepackage{braket}

\makeatother

\begin{document}
\begin{figure}
\global\long\def\del{\boldsymbol{\nabla}}
\global\long\def\x{\times}
\global\long\def\t{\cdot}
\global\long\def\d{\mathrm{d}}
\global\long\def\ket#1{\left|#1\right\rangle }
\global\long\def\bra#1{\left\langle #1\right|}
\global\long\def\braket#1#2{\langle#1|#2\rangle}
\global\long\def\da{\dagger}
\global\long\def\avg#1{\left\langle #1\right\rangle }
\global\long\def\h{\hbar}
\global\long\def\eval#1{\left.#1\right|}
\global\long\def\b#1{\boldsymbol{#1}}
\global\long\def\txt#1{\mathrm{#1}}
\global\long\def\i{\mathrm{i}}
\global\long\def\e{\mathrm{e}}
\global\long\def\ra{\rightarrow}
\end{figure}

\title{Propensity rules for photoelectron circular dichroism in strong field ionization of chiral molecules}

\author{Andres F. Ordonez}
\email{andres.ordonez@icfo.eu}
\affiliation{Max-Born-Institut, Berlin, Germany}
\thanks{Present address: ICFO - Institut de Ciencies Fotoniques, Barcelona, Spain}
\author{Olga Smirnova}
\email{olga.smirnova@mbi-berlin.de}
\affiliation{Max-Born-Institut, Berlin, Germany}
\affiliation{Technische Universit\"at Berlin, Berlin, Germany}

\begin{abstract}
Chiral molecules ionized by circularly polarized fields produce a photoelectron current orthogonal to the polarization plane.
This current has opposite directions for opposite enantiomers 
and provides an extremely sensitive probe of molecular handedness.
Recently, such photoelectron currents have been measured in the strong-field ionization regime, where they may serve as an ultrafast probe of molecular chirality.
Here we provide a mechanism for the emergence of such strong-field photoelectron currents in terms of two propensity rules that link the properties of the initial electronic chiral state to the direction of the photoelectron current. 
\end{abstract}
\maketitle

\section{Introduction}

Molecular chirality  plays a key role in the operation of living organisms,  
production of drugs, fragrances, agrochemicals,  and molecular
machines \cite{koumura_light-driven_1999}. Thus, creating new schemes for efficient  chiral discrimination and enantioseparation is important from fundamental and practical standpoints. 

The photoionization of an isotropic ensemble of chiral molecules with circularly polarized light belongs to a new set of methods discriminating molecular enantiomers without the help of the magnetic component of the light field and therefore does it in a new and  extremely efficient way \cite{ordonez_generalized_2018}. Photoionization  causes a pronounced forward-backward asymmetry
(FBA) in the photoelectron angular distribution (PAD) \cite{ritchie_theory_1976,powis00,bowering_asymmetry_2001}  that depends
on the relative handedness between the sample and the ``electric field
+ detector'' system \cite{ ordonez_generalized_2018}. This phenomenon, known
as photoelectron circular dichroism (PECD), has been the subject of
an increasing number of investigations in the one- and few-photon
regimes \cite{lux_circular_2012,nahon_valence_2015,dreissigacker_photoelectron_2014, demekhin_photoelectron_2018}, and very
recently it was also observed in the many-photon tunneling regime
\cite{beaulieu_universality_2016,fehre_link_2019}.


In a previous work \cite{ordonez_propensity_2019} we introduced three families of chiral
wave functions, classified according to the origin of their chirality, and built from hydrogenic
wave functions.
We used these wave functions to understand
how the chirality of the initial state can lead to 
PECD in the one-photon case in a simplified setting where the  continuum states are isotropic and the molecules are aligned perpendicular to the polarization plane. We found that in this case PECD emerges 
as the result
of two simple propensity rules that explicitly connect the circular
motion of the electron in the plane of the circularly polarized field
with the linear motion perpendicular to the plane, responsible for
the FBA. Now we turn our attention to the understanding of PECD in
the many-photon ionization regime by taking advantage of the atomic
nature of the chiral hydrogenic wave functions, which is ideally suited
to include the effect of chirality in the PPT analytical theory of
strong field ionization \cite{perelomov_ionization_1966}. As in our previous work, we will approach this problem in a simplified setting where: (i) the molecules are assumed to be aligned perpendicular to the polarization plane and (ii) the effect of the anisotropy of the molecular potential on the photoelectron is neglected. Assumption (i) is experimentally achievable and assumption (ii) is reasonable in the tunneling picture, where the electron exits the tunnel far from the parent ion\footnote{The anisotropy of the molecular potential is encoded in multipole terms higher than the monopole which decay rapidly with increasing distance to the parent ion.}. 

\section{Physical picture\label{sec:Physical-picture}}

We will study the photoionization produced by the interaction between
a circularly polarized ($\sigma=\pm1$ ) electric field of amplitude $\mathcal{E}$ and frequency $\omega$, propagating along
the $z$ axis
, 
\begin{equation}
\vec{E}_{\sigma}\left(t\right)=\mathcal{E}\left[\cos\left(\omega t\right)\hat{x}+\sigma\sin\left(\omega t\right)\hat{y}\right],\label{eq:E}
\end{equation}
and a chiral hydrogen atom \cite{ordonez_propensity_2019} in the initial state
\begin{eqnarray}
\chi_{\rho}^{\epsilon}\left(\vec{r}\right) & \equiv & \frac{1}{\sqrt{2}}\left[\chi_{\mathrm{c}}^{\epsilon}\left(\vec{r}\right)+\chi_{\mathrm{c}}^{\epsilon*}\left(\vec{r}\right)\right],\qquad\epsilon=\pm,\label{eq:chi_rho}
\end{eqnarray}
where
\begin{equation}
\chi_{\mathrm{c}}^{\pm}\left(\vec{r}\right)\equiv\frac{1}{\sqrt{2}}\left[\psi_{4,2,\pm1}\left(\vec{r}\right)+\i\psi_{4,3,\pm1}\left(\vec{r}\right)\right],\label{eq:chi_c}
\end{equation}
\begin{equation}
\chi_{\mathrm{c}}^{\pm*}\left(\vec{r}\right)=\frac{1}{\sqrt{2}}\left[-\psi_{4,2,\mp1}\left(\vec{r}\right)+\i\psi_{4,3,\mp1}\left(\vec{r}\right)\right].\label{eq:chi_c_cc}
\end{equation}
Here the superscript $\epsilon=\pm$ indicates the handedness of chiral states,  $\psi_{n,l,m}$ denotes the hydrogenic
state with principal quantum number $n,$ angular momentum quantum
number $l$, and magnetic quantum number $m$. Equation \eqref{eq:chi_c_cc}
follows from Eq. \eqref{eq:chi_c} and the property of spherical harmonics
$Y_{l}^{m*}=\left(-1\right)^{m}Y_{l}^{-m}$. The states $\chi_{\rho}^{\epsilon}$
and $\chi_{\mathrm{c}}^{\epsilon}$ are instances of the chiral-density
and chiral-current families of hydrogenic chiral states introduced
in Ref. \cite{ordonez_propensity_2019}, respectively. The  superscript $\epsilon=\pm$ indicating the
enantiomer simply corresponds to the sign of $m$, as can be seen in Eqs. \eqref{eq:chi_rho}-\eqref{eq:chi_c_cc}. Opposite enantiomers ($+$
and $-$) are related to each other through a reflection in the $x=0$
plane, which by definition is equivalent to a reversal of the sign
of $m$ used in the corresponding hydrogenic wave functions. 

\begin{figure*}
\noindent \begin{centering}
\includegraphics[width=0.9\textwidth]{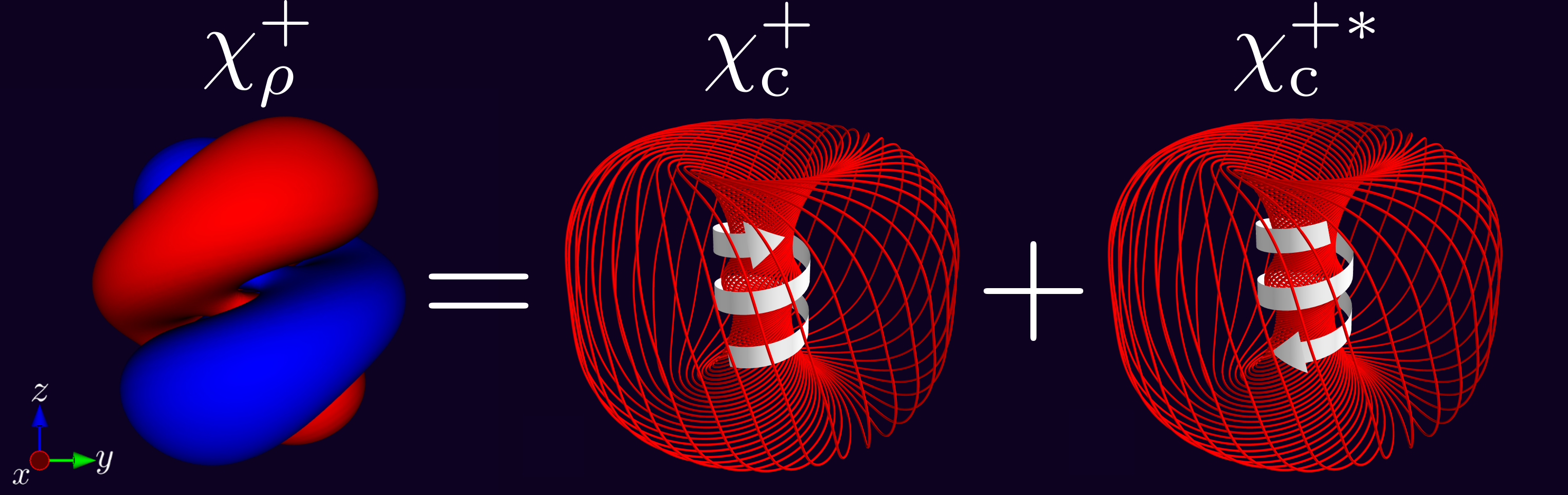}
\par\end{centering}
\caption{Sketch of the decomposition of the \emph{chiral density state} $\chi_{\rho}^{+}$
into\emph{ }the \emph{chiral current states} $\chi_{\mathrm{c}}^{+}$
and $\chi_{\mathrm{c}}^{+*}$, see Eq. \eqref{eq:chi_rho} and
Ref. \cite{ordonez_propensity_2019}. Left: Isosurfaces $\vert\chi_{\mathrm{\rho}}^{+}\left(\vec{r}\right)\vert=\pm0.001\,\mathrm{a.u.}$
Right: Trajectory followed by an element of the probability fluid $\left|\chi_{\mathrm{c}}^{+}\left(\vec{r}\right)\right|^{2}$
for the states $\chi_{\mathrm{c}}^{+}$ and $\chi_{\mathrm{c}}^{+*}$.
The white arrows indicate the direction of the flow.
\label{fig:chi_rho_decomposition}}
\end{figure*}

Although both $\chi_{\rho}^{\epsilon}$ and $\chi_{\mathrm{c}}^{\epsilon}$
display chirality, it manifests itself differently in each state.
As can be seen in Fig. \ref{fig:chi_rho_decomposition} for $\epsilon=+$,
the chirality of $\chi_{\rho}^{\epsilon}$ is encoded in its helical
probability density $\left|\chi_{\rho}^{\epsilon}\left(\vec{r}\right)\right|^{2}$,
while the chirality of $\chi_{\mathrm{c}}^{\epsilon}$ is encoded
in its torus-knot-like probability current $\vec{j}\left(\vec{r};\chi_{\mathrm{c}}^{\epsilon}\right)$,
which is visualized in Fig. \ref{fig:chi_rho_decomposition} via the
trajectory followed by an element of the probability fluid $\left|\chi_{\mathrm{c}}^{\epsilon}\left(\vec{r}\right)\right|^{2}$.
In analogy to how a standing plane wave can be decomposed into two
plane waves traveling in opposite directions, Eq. \eqref{eq:chi_rho}
shows how the\emph{ chiral density state} $\chi_{\rho}^{\epsilon}\left(\vec{r}\right)$,
which corresponds to a real function and therefore has no probability
current, can be decomposed into two \emph{chiral current states}
$\chi_{\mathrm{c}}^{\epsilon}\left(\vec{r}\right)$ and $\chi_{\mathrm{c}}^{\epsilon*}\left(\vec{r}\right)$,
with opposite probability currents $\vec{j}\left(\vec{r};\chi_{\mathrm{c}}^{\epsilon}\right)$
and $\vec{j}\left(\vec{r};\chi_{\mathrm{c}}^{\epsilon*}\right)=-\vec{j}\left(\vec{r};\chi_{\mathrm{c}}^{\epsilon}\right)$.

Both in the one- and in the many-photon regimes, the photoionization
yield depends on the relative sense of rotation between the circularly polarized
electric field and the bound electronic current in the plane of polarization.
In the one-photon regime the total photoionization yield is greater
when the bound electron and the field rotate in the same direction
\cite{hans_a._bethe_quantum_1957}, while in the many-photon regime
it is greater when the electron and the field rotate in opposite directions
\cite{barth_nonadiabatic_2011,herath_strong-field_2012,barth_nonadiabatic_2013,eckart2018ultrafast,Beiser_PRA_2004,Bergues_PRL_2005}.
We shall call this dependence \emph{propensity rule 1} (PR1). Furthermore,
we have shown in Ref. \cite{ordonez_propensity_2019} that in the one-photon case, the
component of the bound electronic current perpendicular to the plane
of polarization in the region close to the core $j_{z}\left(\vec{r}\ra0;\chi_{\mathrm{c}}^{\epsilon}\right)$
is projected onto the continuum by the ionizing photon, and gives
rise to an excess of photoelectrons either in the forward $\left(+z\right)$
or backward ($-z$) direction
. We shall call this dependence \emph{propensity rule 2} (PR2). Therefore,
even though in the state $\chi_{\rho}^{\epsilon}$ the electron currents
of $\chi_{\mathrm{c}}^{\epsilon}$ and $\chi_{\mathrm{c}}^{\epsilon*}$
cancel each other, PR1 determines which state, $\chi_{\mathrm{c}}^{\epsilon}$
or $\chi_{\mathrm{c}}^{\epsilon*}$, dominates the photoelectron spectrum,
and PR2 applied to the dominant state determines whether more electrons
go forwards or backwards. 
As mentioned above, we know that PR1 is reversed when going from the
one- to the many-photon regime, and we know the form of PR2 in the
one-photon regime. In the many-photon regime, the adiabatic tunneling
picture suggests that the photoelectron current along $z$ will reflect
that of the bound wave function under the barrier and in the vicinity of the tunnel exit, because of the
continuity of the wave function $\chi_{\mathrm{c}}^{\epsilon}\left(\vec{r}\right)$
and its derivatives $\vec{\nabla}\chi_{\mathrm{c}}^{\epsilon}\left(\vec{r}\right)$
across the exit of the tunnel. 
Since
$j_{z}\left(\vec{r};\chi_{\mathrm{c}}^{\epsilon}\right)$ has opposite
signs close to and far from the core and the tunnel exit is far from
the core, this means that PR2 will also be reversed when going from
the one- to the many-photon regime. The simultaneous reversal of PR1
and PR2 when passing from the one- to the many-photon regime means
that overall, the FBA resulting from photoionization of $\chi_{\rho}^{\epsilon}$
will have the same sign in both regimes, that is, if more photoelectrons
are ejected forward (backward) for an initial state $\chi_{\rho}^{\epsilon}$
and a given polarization of the electric field in the one-photon regime,
this will also be the case in the many-photon regime. Below we
apply the PPT theory of strong-field
ionization to chiral hydrogen to prove the physical picture described in this section.

\section{Theory}
\subsection{Strong field ionization of atomic states}
Following the PPT theory \cite{perelomov_ionization_1966,perelomov_ionization_1967-1,barth_nonadiabatic_2013}, one can show that the cycle-averaged
current probability density asymptotically far from the nucleus resulting
from strong field ionization of an atom in an initial state $\psi_{l,m}$ via a long and circularly
polarized pulse [Eq. (\ref{eq:E})] can be expressed as a sum over multiphoton channels
according to 
\begin{equation}
\vec{j}_{\sigma}\left(\vec{r};\psi_{l,m}\right)=\frac{1}{r^{2}}\sum_{n=n_{0}}^{\infty}W_\sigma(\vec{k}_n;\psi_{l,m})\vec{k}_{n},\label{eq:j}
\end{equation}
where $\vec{k}_{n}$ is the photoelectron momentum
measured at the detector, it is parallel to $\vec{r}$ and its magnitude satisfies
\begin{equation}
\frac{k_{n}^{2}}{2}=n\omega-2U_{p}-I_{p},\label{eq:energy_conservation}
\end{equation}
$n$ is the number of absorbed photons, $2U_{p}=A_{0}^{2}/2$ is the
average kinetic energy of an electron in the circularly polarized
electric field \eqref{eq:E}, $A_{0}=\mathcal{E}/\omega$ is the amplitude
of the vector potential, $I_{p}$ is the ionization potential, and
$n_{0}$ is the minimum number of photons required for ionization
in a strong field. $W_\sigma(\vec{k}_n;\psi_{l,m})$ is the probability of populating 
a Volkov
state with drift momentum $\vec{k}_{n}$, i.e. it is the PAD at the energy
$k_{n}^{2}/2$ \cite{barth_nonadiabatic_2013} 
\begin{equation}
W_\sigma(\vec{k}_n;\psi_{l,m})\equiv\left|\left[\frac{v_{n\sigma}^{2}\left(t_{i}\right)}{2}+I_{p}\right]\tilde{\psi}_{l,m}\left(\vec{v}_{n\sigma}\left(t_{i}\right)\right)\right|^{2}G_n(k_z),\label{eq:PAD}
\end{equation}
where we defined\footnote{In comparison to the notation used in Ref. \cite{barth_nonadiabatic_2013},
we did the replacement $\chi_{n}\left(k_{z}\right)\ra X_{n}\left(k_{z}\right)$
in order to avoid confusion with the symbols $\chi_{\mathrm{c}}^{\epsilon}$
and $\chi_{\rho}^{\epsilon}$ that we use here for the chiral wave
functions. }
\begin{equation}
    G_n(k_z) \equiv \frac{\e^{-2n\left[\cosh^{-1}X_{n}\left(k_{z}\right)-\sqrt{1-1/X_{n}^{2}\left(k_{z}\right)}\right]}}{ n\sqrt{1-1/X_{n}^{2}\left(k_{z}\right)}},
    \label{eq:G_n}
\end{equation}
\begin{equation}
X_{n}\left(k_{z}\right)\equiv\frac{n\omega}{A_{0}\sqrt{k_{n}^{2}-k_{z}^{2}}},
\end{equation}
$\tilde{\psi}_{lm}\left(\vec{v}\right)$ is the wave function of
the initial state in the momentum representation, 
\begin{equation}
\tilde{\psi}_{lm}\left(\vec{v}\right)=\frac{1}{\left(2\pi\right)^{3/2}}\int\mathrm{d}\vec{r}\,\e^{-\i\vec{v}\t\vec{r}}\psi_{l,m}\left(\vec{r}\right),
\end{equation}
and $\vec{v}_{n\sigma}\left(t_{i}\right)\equiv\vec{k}_{n}+\vec{A}_{\sigma}\left(t_{i}\right)$
is the velocity of the electron, which depends on the vector potential
$\vec{A}_{\sigma}\left(t_{i}\right)$ at the complex time $t_{i}$\footnote{The emergence of a complex time $t_{i}$ in the theory results from
the use of the saddle point approximation for the calculation of a
time integral.}. The latter is defined through the saddle point equation $v_{n\sigma}^{2}\left(t_{i}\right)=-2I_{p}$,
and corresponds to the time at which the electron enters the potential
barrier that results from the bending of the binding potential by
the strong electric field \cite{perelomov_ionization_1967-1}.  
Despite the saddle point equation $v_{n\sigma}^{2}\left(t_{i}\right)=-2I_{p}$,
Eq. \eqref{eq:PAD} does not vanish because $\vec{v}_{n}\left(t_{i}\right)$
is a pole of $\tilde{\psi}_{lm}\left(\vec{v}\right)$. Furthermore,
since the behavior of the wave function in momentum space $\tilde{\psi}_{lm}\left(\vec{v}\right)$
close to a pole $v_{\sigma}\left(t_{i}\right)$ only contains information
about the asymptotic part of its counterpart in coordinate space $\psi_{l,m}\left(\vec{r}\right)$ \cite{perelomov_ionization_1966,gribakin1997multiphoton},
the latter can be replaced by its asymptotic form, which for a spherically-symmetric short-range
potential is given by
\begin{equation}
\lim_{r\ra\infty}\psi_{lm}\left(\vec{r}\right)=C_{\kappa,l}\kappa^{3/2}\frac{\e^{-\kappa r}}{\kappa r}Y_{l}^{m}\left(\hat{r}\right),\label{eq:asymptotic}
\end{equation}
and where the constant $C_{\kappa,l}$ contains the information about
the short-range behavior of $\psi_{l,m}\left(\vec{r}\right)$.
 Using Eq. \eqref{eq:asymptotic} one can show \cite{barth_nonadiabatic_2013} that 
the fingerprint of the initial state on the PAD {[}Eq. \eqref{eq:PAD}{]}
reduces to 
\begin{equation}
\left[\frac{v_{n\sigma}^{2}\left(t_{i}\right)}{2}+I_{p}\right]\tilde{\psi}_{lm}\left(\vec{v}_{n\sigma}\left(t_{i}\right)\right)=C_{\kappa,l}\sqrt{\frac{\kappa}{2\pi}}\left(\frac{v_{n\sigma}\left(t_{i}\right)}{\kappa}\right)^{l}Y_{l}^{m}\left(\hat{v}_{n\sigma}\left(t_{i}\right)\right)\e^{-\i l\pi/2}.\label{eq:orbital_contribution-1}
\end{equation}

\subsection{Strong field ionization of a chiral state}
Up to this point the theory has followed Ref. \cite{barth_nonadiabatic_2013},
which assumes an initial state $\psi_{l,m}$ with well defined angular momentum
quantum numbers $\left(l,m\right)$, and therefore a central potential.
To obtain enantio-sensitive current triggered by strong field ionization, we will replace
the initial state $\psi_{l,m}\left(\vec{r}\right)$ in the derivation above by a chiral hydrogen state, where 
\begin{equation}
C_{n,l}=\frac{\left(-1\right)^{n-l-1}2^{n}}{\sqrt{n\left(n+l\right)!\left(n-l-1\right)!}}.
\end{equation}
Replacing $\psi_{l,m}\left(\vec{r}\right)$ by $\chi_{\mathrm{c}}^{\epsilon}\left(\vec{r}\right)$,
and using the corresponding asymptotic expression \eqref{eq:asymptotic}
for each partial wave in $\chi_{\mathrm{c}}^{\epsilon}\left(\vec{r}\right)$
yields  (see Appendix) 
\begin{eqnarray}
W_\sigma(\vec{k}_n;\chi_{\mathrm{c}}^{\epsilon}) =  \left[A\left(k_{z}\right)+B\left(k_{z}\right)\right] D(k_z) \left|\e^{\i\epsilon\varphi_{v\sigma}\left(t_{i}\right)}\right|^{2}G_n(k_z),\label{eq:PAD_orbital_chi}
\end{eqnarray}
where $A\left(k_{z}\right)$ and $D(k_z)$ are even  polynomials of $k_{z}$ while $B\left(k_{z}\right)$ is an odd polynomial of $k_{z}$,
\begin{equation}
A\left(k_{z}\right)\equiv\bigg[175\left(\frac{k_{z}}{\kappa}\right)^{4}C_{\kappa,3}^{2}+10\left(\frac{k_{z}}{\kappa}\right)^{2}\left(4C_{\kappa,2}^{2}+7C_{\kappa,3}^{2}\right)+7C_{\kappa,3}^{2}\bigg],\label{eq:A}
\end{equation}
\begin{equation}
B\left(k_{z}\right)\equiv4\sqrt{70}C_{\kappa,2}C_{\kappa,3}\bigg[5\left(\frac{k_{z}}{\kappa}\right)^{2}+1\bigg]\frac{k_{z}}{\kappa},\label{eq:B}
\end{equation}
\begin{equation}
D\left(k_{z}\right)\equiv\frac{3\kappa}{2^{8}\pi^{2}}\left[\left(\frac{k_{z}}{\kappa}\right)^{2}+1\right].\label{eq:D}
\end{equation}
The factor $\left|\e^{\i\epsilon\varphi_{v\sigma}\left(t_{i}\right)}\right|^{2}$,
which is not equal to unity because the so-called tunneling-momentum
angle $\varphi_{v\sigma}\left(t_{i}\right)$ is complex, gives rise
to PR1, and it is given by (see \cite{barth_nonadiabatic_2011,barth_nonadiabatic_2013})
\begin{equation}
\left|\e^{\i\epsilon\varphi_{v\sigma}\left(t_{i}\right)}\right|^{2}=\frac{I_{p}\left\{ 2X_{n}^{2}\left(k_{z}\right)\left[1-\sigma\epsilon\sqrt{1-1/X_{n}^{2}\left(k_{z}\right)}\right]-\left(1+\gamma^{2}\right)n/n_{0}\right\} ^{2}}{2\gamma^{2}X_{n}^{2}\left(k_{z}\right)\left(k_{z}^{2}+2I_{p}\right)},
\end{equation}
where $\gamma=\sqrt{2I_{p}}/A_{0}$ is the Keldysh parameter \cite{keldysh_ionization_1965}.
As expected from symmetry, this term behaves so that a reversal of
the polarization $\sigma$ is equivalent to a reversal
of the azimuthal probability current of the bound state in the polarization
plane $\mathrm{sgn}\,m=\epsilon$, i.e. 
\begin{equation}
\left|\e^{\i\epsilon\varphi_{v,-\sigma}\left(t_{i}\right)}\right|^{2}=\left|\e^{-\i\epsilon\varphi_{v,\sigma}\left(t_{i}\right)}\right|^{2}.\label{eq:symmetry}
\end{equation}
In other words, the angle-integrated yield is only affected by the
relative direction of the probability current of the bound state in
the polarization plane with respect to the direction of rotation of
the electric field. Furthermore, since in the case we are considering,
opposite values of $m$ correspond to opposite enantiomers, we
have that for the $\chi_{\mathrm{c}}^{\epsilon}$ states, opposite
enantiomers subject to opposite polarizations display the same angle-integrated
yield.

Since $A(k_z)$, $D(k_z)$, $|\e^{\i\epsilon\varphi_{v,\sigma}(t_{i})}|^{2}$, and $G_n(k_z)$ are even functions of $k_z$, Eq. (\ref{eq:PAD_orbital_chi}) shows that 
the FBA is entirely encoded in the odd polynomial $B\left(k_{z}\right)$.
Furthermore, since $\mathrm{sgn}\left(C_{\kappa,l+1}C_{\kappa,l}\right)=-1$\footnote{This follows from considering that the number of zeros of the radial
part of the wave function is given by $n-l-1$ and the convention
of setting the radial wave function to be positive as $r\ra0$. One
can of course also use a different convention, but then the relative
phases between the hydrogenic states in Eqs. \eqref{eq:chi_rho} and
\eqref{eq:chi_c} also have to be modified accordingly to keep the
same density and probability currents discussed before. Our conclusions
are independent from the convention. }, we can see from the expression for $B\left(k_{z}\right)$ and from
Eqs. \eqref{eq:PAD} and \eqref{eq:PAD_orbital_chi} that more photoelectrons
will be emitted backwards ($-z$) than forwards ($+z$), for either
polarization $\sigma=\pm1$ of the electric field.

From the expressions for $\chi_{\mathrm{c}}^{\epsilon}$ and $\chi_{\mathrm{c}}^{\epsilon*}$
{[}Eqs. \eqref{eq:chi_c} and \eqref{eq:chi_c_cc}{]} and from Eqs. \eqref{eq:PAD_orbital_chi}-\eqref{eq:B}, it
follows that the PAD for the complex conjugated state $\chi_{\mathrm{c}}^{\epsilon*}$ reads 
\begin{equation}
W_\sigma(\vec{k}_n;\chi_{\mathrm{c}}^{\epsilon*})=\left[A\left(k_{z}\right)-B\left(k_{z}\right)\right]D(k_z)\left|\e^{-\i\epsilon\varphi_{v\sigma}\left(t_{i}\right)}\right|^{2}G_n(k_z),\label{eq:PAD_orbital_chi*}
\end{equation}
which differs from the corresponding equation for $\chi_c^\epsilon$ [Eq. (\ref{eq:PAD_orbital_chi})] in the signs in front of $B(k_z)$ and $\epsilon$. This shows that $\chi_{\mathrm{c}}^{\epsilon*}$ yields a FBA exactly
opposite to that of $\chi_{\mathrm{c}}^{\epsilon}$.

Equations \eqref{eq:PAD_orbital_chi}
and \eqref{eq:PAD_orbital_chi*} yield the following important conclusions:

First, they confirm our expectation that the
FBA is determined by the direction of the bound probability current
close to the tunnel exit. 

Second, the product of $C_{\kappa,2}$ and $C_{\kappa,3}$ in Eq. \eqref{eq:B} shows
that the FBA emerges exclusively from the interference between the
two components, $\psi_{4,2,\pm1}$ and $i\psi_{4,3,\pm1}$, that make up the $\chi_{\mathrm{c}}^{\epsilon}$ state.  
As can be seen in the Appendix, 
this interference
vanishes when the relative phase $\e^{\i\eta}$ between the two components
is $\pm\pi$.  
That is, the chiral states $\chi_{\mathrm{p}}^{\pm}\equiv(\psi_{4,2,\pm1} + \psi_{4,2,\pm1})/\sqrt{2}$ introduced in Ref. \cite{ordonez_propensity_2019}, which 
instead of a probability
current along $z$ have probability density polarized along $z$ (see Fig. 1 in Ref. \cite{ordonez_propensity_2019}),
do not display any FBA in the case of strong field of ionization.

For the state $\chi_{\rho}^{\epsilon}$, which has a chiral probability density and can be decomposed into states $\chi_{\mathrm{c}}^{\epsilon}$ and $\chi_{\mathrm{c}}^{\epsilon*}$,
one can show (see Appendix) 
that the PAD at energy $k_n^2/2$, averaged over the contributions of all initial state orientations related
to the original orientation {[}Eq. \eqref{eq:chi_rho}{]} by a rotation
$R_z(\alpha)$ of $\alpha$ radians around the $z$ axis (as would be appropriate
if the state is perfectly aligned along the $z$ axis\footnote{Note that the state $\chi_\rho^\epsilon(\vec{r})$ is symmetric with respect to rotations by $\pi$ around the $y$ axis.}), is given by the sum of the PADs for the states $\chi_{\mathrm{c}}^{\epsilon}$ and $\chi_{\mathrm{c}}^{\epsilon*}$
{[}Eqs. \eqref{eq:PAD_orbital_chi} and \eqref{eq:PAD_orbital_chi*}{]},
\begin{align}
\overline{W}_{\sigma}(\vec{k}_n;\chi_{\rho}^{\epsilon}) \equiv 
\frac{1}{2\pi} & \int_{0}^{2\pi}\mathrm{d}\alpha\, W_{\sigma}(\vec{k}_n;R_z(\alpha)\chi_{\rho}^{\epsilon}) \nonumber \\ 
= \frac{1}{2}\bigg[ & W_{\sigma}(\vec{k}_n;\chi_{\mathrm{c}}^{\epsilon}) + W_{\sigma}(\vec{k}_n;\chi_{\mathrm{c}}^{\epsilon*}) \bigg] \nonumber \\
= \frac{1}{2}\bigg\{ & A(k_{z})\left[\left|\e^{\i\epsilon\varphi_{v\sigma}\left(t_{i}\right)}\right|^{2}+\left|\e^{-\i\epsilon\varphi_{v\sigma}\left(t_{i}\right)}\right|^{2}\right]\nonumber \\
+ & B(k_{z})\left[\left|\e^{\i\epsilon\varphi_{v\sigma}\left(t_{i}\right)}\right|^{2}-\left|\e^{-\i\epsilon\varphi_{v\sigma}\left(t_{i}\right)}\right|^{2}\right]\bigg\}D(k_z)G_n(k_z),
 \label{eq:PAD_orbital_chi_rho}
\end{align}

Equation \eqref{eq:PAD_orbital_chi_rho} 
clearly shows that the asymmetric
response along $z$ encoded in $B\left(k_{z}\right)$ is coupled to
the dichroic and enantio-sensitive response encoded in the difference
$\vert\e^{\i\epsilon\varphi_{v\sigma}\left(t_{i}\right)}\vert^{2}-\vert\e^{-\i\epsilon\varphi_{v\sigma}\left(t_{i}\right)}\vert^{2}$,
so that either opposite enantiomers (opposite values of
$\epsilon$) or opposite circular polarizations (opposite values of
$\sigma$) yield opposite FBAs [see Eq. \eqref{eq:symmetry}]. That is, the enantiosensitive and dichroic response is encoded entirely
in the second term of Eq. \eqref{eq:PAD_orbital_chi_rho}. Furthermore,
unlike $\chi_{\mathrm{c}}^{\epsilon}$ and $\chi_{\mathrm{c}}^{\epsilon*}$,
which have different angle-integrated yields for either opposite enantiomers
or opposite circular polarizations, the angle-integrated photoelectron
yield for the initial state $\chi_{\rho}^{\epsilon}$ is independent
of the enantiomer and circular polarization used. This is because the contribution from the second term in Eq. (\ref{eq:PAD_orbital_chi_rho}) to the angle integrated yield vanishes and the first term in Eq. (\ref{eq:PAD_orbital_chi_rho}) is explicitly symmetric with respect to to a reversal of either polarization or enantiomer.

Using Eq. \eqref{eq:symmetry} we get that the ratio of the dichroic and non-dichroic responses, which is equivalent to the ratio of  enantiosensitive and non-enantiosensitive responses, is given by
\begin{align}
    \frac{\overline{W}_{\sigma}(\vec{k}_n;\chi_{\rho}^{\epsilon}) - \overline{W}_{-\sigma}(\vec{k}_n;\chi_{\rho}^{\epsilon})}{\overline{W}_{\sigma}(\vec{k}_n;\chi_{\rho}^{\epsilon}) + \overline{W}_{-\sigma}(\vec{k}_n;\chi_{\rho}^{\epsilon})} & = \frac{B(k_{z})}{A(k_{z})}\frac{\left|\e^{\i\epsilon\varphi_{v\sigma}\left(t_{i}\right)}\right|^{2}-\left|\e^{-\i\epsilon\varphi_{v\sigma}\left(t_{i}\right)}\right|^{2}}{\left|\e^{\i\epsilon\varphi_{v\sigma}\left(t_{i}\right)}\right|^{2}+\left|\e^{-\i\epsilon\varphi_{v\sigma}\left(t_{i}\right)}\right|^{2}},\nonumber\\
    & = \frac{B(k_{z})}{A(k_{z})} \frac{1-R(k_z)}{1+R(k_z)},
\end{align}
where $R\equiv|\e^{-\i\epsilon\varphi_{v\sigma}(t_{i})}|^{2}/|\e^{\i\epsilon\varphi_{v\sigma}(t_{i})}|^{2}$ is the ratio of ionization rates for co- and counter-rotating electrons (see Eq. (100) of \cite{Kaushal_PRA_2013}).

\section{Calculations}

In view of the results obtained in Refs. \cite{ordonez_generalized_2018, ordonez_propensity_2019,ordonez_propensity_2019_2}, we will base our analysis on the photoelectron current\footnote{We will use the term ``current'' as a shorthand for ``probability
current density'' and we will omit the $1/r^{2}$ scaling term
in Eq. \eqref{eq:j}.} $\vec{j}(\vec{k}_{n};\psi)$ associated with the photoelectron of momentum
$\vec{k}_{n}$ and the initial state $\psi$,
\begin{equation}
\vec{j}_{\sigma}(\vec{k}_{n};\psi)\equiv W_{\sigma}(\vec{k}_{n};\psi)\vec{k}_{n}.
\end{equation}
For a given $n$-photon channel, the net photoelectron current along
$z$ reads
\begin{eqnarray}
j_{\sigma,z}\left(k_{n};\psi\right) =  \int\mathrm{d}\Omega_{k}\,j_{\sigma,z}(\vec{k}_{n};\psi)
 =  \int\mathrm{d}\Omega_{k}\,W_\sigma(\vec{k}_n;\psi)k_{n,z}\label{eq:jz}
\end{eqnarray}
where the integration is over directions of $\vec{k}_{n}$. Due
to the symmetry of the system, $j_{\sigma,z}(\vec{k}_n)$ is the only non-zero Cartesian
component of $\vec{j}_{\sigma}(\vec{k}_n)$. 
The angle-integrated radial component of the photoelectron
current is also of interest as it determines the total ionization yield. It is given by
\begin{eqnarray}
j_{\sigma,r}\left(k_{n};\psi\right) = \int\mathrm{d}\Omega_{k}\,j_{\sigma,r}(\vec{k};\psi)
 = \int\mathrm{d}\Omega_{k}\,W_{\sigma}(\vec{k}_n;\psi)k_{n}.\label{eq:jr}
\end{eqnarray}
As shown in Refs. \cite{ordonez_generalized_2018, ordonez_propensity_2019} and as can be seen from
Eqs. \eqref{eq:jz} and \eqref{eq:jr}, if one expands the PAD in
Legendre polynomials, 
\begin{equation}
W_{\sigma}(\vec{k}_n;\psi)=\sum_{l=0}^{\infty}b_{l}^{(\sigma)}\left(k_{n};\psi\right)P_{l}\left(\cos\theta_{k}\right)
\end{equation}
it becomes clear that the radial and $z$ components of the current are proportional to the zeroth and first order coefficients, respectively, 
\begin{equation}
j_{\sigma,r}\left(k_{n};\psi\right)=4\pi k_n b_{0}^{(\sigma)}\left(k_{n};\psi\right),\qquad j_{\sigma,z}\left(k_{n};\psi\right)=\frac{4\pi}{3}k_n b_{1}^{(\sigma)}\left(k_{n};\psi\right).
\end{equation}
\begin{figure}
\noindent \begin{centering}
\includegraphics[scale=0.35]{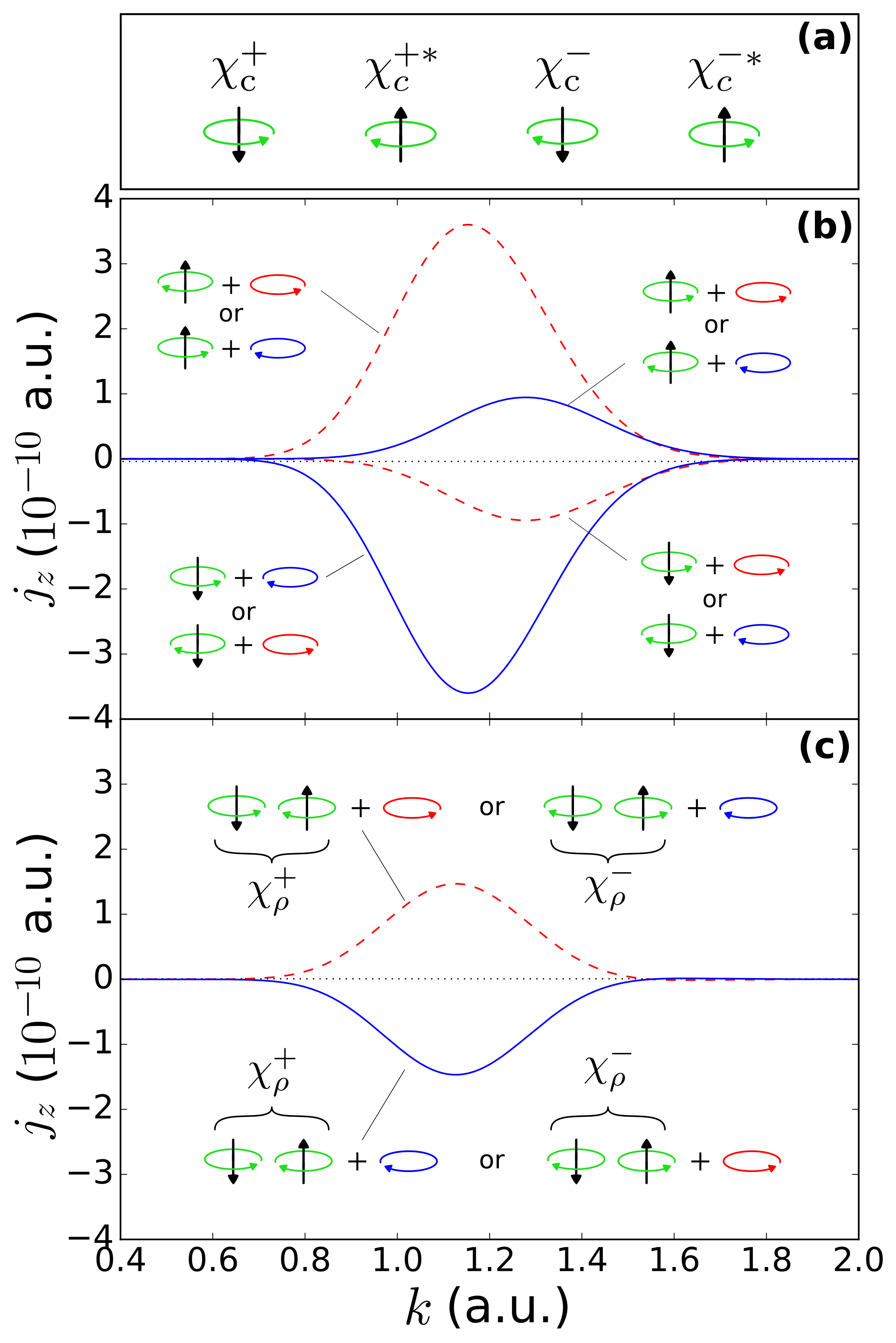}
\par\end{centering}
\caption{Photoelectron current density along $z$ as a function of photoelectron
momentum {[}Eq. \eqref{eq:jz}{]}, resulting from strong field ionization
of the the initial states $\chi_{\mathrm{c}}^{\epsilon}$, $\chi_{\mathrm{c}}^{\epsilon*}$,
and $\chi_{\rho}^{\epsilon}$ {[}Eqs. \eqref{eq:chi_rho}-\eqref{eq:chi_c_cc}
and Fig. \ref{fig:chi_rho_decomposition}{]}, via intense light circularly
polarized in the $xy$ plane. \textbf{(a)} Diagrams indicating the
directions of the azimuthal component (green circular arrow) and the
vertical component of the probability current in the region close
to the tunnel exit (black vertical arrow) in the bound states $\chi_{\mathrm{c}}^{\epsilon}$.
\textbf{(b)} Photoelectron current {[}Eqs. \eqref{eq:PAD}, \eqref{eq:PAD_orbital_chi},
\eqref{eq:PAD_orbital_chi*}, and \eqref{eq:jz}{]} for different combinations of 
initial state {[}indicated according to (a){]} and light polarization
(red or blue circular arrow after the plus sign). Note
that the direction of $j_{z}$ is determined by the direction of the
vertical component of the bound current in the region close to the
tunnel exit. Except for the high-momentum tail beyond $k\approx1.5\,\mathrm{a.u.}$,
the magnitude of $j_{z}$ is greater when the azimuthal bound probability
current and the ionizing light rotate in opposite directions. \textbf{(c)}
Same as (b) but for the initial state $\chi_{\rho}^{\epsilon}$ after
averaging over the orientations of the initial state that result from
a rotation around the $z$ axis {[}Eqs. \eqref{eq:PAD}, \eqref{eq:PAD_orbital_chi_rho},
and \eqref{eq:jz}{]}. The state $\chi_{\rho}^{\epsilon}$ is indicated
with the two consecutive diagrams corresponding to its decomposition
into states $\chi_{\mathrm{c}}^{\epsilon}$ and $\chi_{\mathrm{c}}^{\epsilon*}$.
In this case, $j_{z}$ is the average of the results obtained for
each of its components in (b) {[}see Eq. \eqref{eq:PAD_orbital_chi_rho}{]}.
The results shown are for a field of amplitude $\mathcal{E}=0.06\,\mathrm{a.u.}$
and frequency $\omega=0.057\,\mathrm{a.u.}$, and for an ionization
potential $I_{p}=0.5\,\mathrm{a.u.}$ \label{fig:j_z_summary}}
\end{figure}
Clearly, only the $k_z$-even part of $W_\sigma$ contributes to $j_{\sigma,r}$ and only the $k_z$-odd part of $W_\sigma$ contributes to $j_{\sigma,z}$. For the states $\chi_\mathrm{c}^{\epsilon}$, $\chi_\mathrm{c}^{\epsilon*}$, and $\chi_{\rho}^{\epsilon}$ this means that only the part of $W_\sigma$ involving $A(k_z)$ contributes to $j_{\sigma,r}$ and only the part of $W_\sigma$ involving $B(k_z)$ contributes to $j_{\sigma,z}$.

Figure \ref{fig:j_z_summary} shows the photoelectron current along
the $z$ axis as a function of the photoelectron momentum at the detector
for all the different enantiomer-polarization configurations involving
the $\chi_{\mathrm{c}}^{\epsilon}$ and $\chi_{\rho}^{\epsilon}$
states, and left- and right circular polarizations in the $xy$ plane,
and for an electric field amplitude $\mathcal{E}=0.06$ a.u. ($I=1.3 \times 10^{14}\,\mathrm{W/cm^2}$), angular frequency $\omega=0.057$ a.u. ($\lambda=800\,\mathrm{nm}$), and ionization potential $I_p=0.5$ a.u., which yield a Keldysh parameter $\gamma=\sqrt{2I_{p}}\omega/\mathcal{E}=0.95$ corresponding to non-adiabiatic tunneling ionization. 
The results in Fig. \ref{fig:j_z_summary}
clearly show how the FBA is governed by the propensity rules discussed
in Sec. \ref{sec:Physical-picture}. Panel (b) shows how for the states
$\chi_{\mathrm{c}}^{\epsilon}$, the direction of the net photoelectron
current coincides with that of the bound probability current in the
region close to the tunnel exit (vertical arrow), that is, 
where $x^{2}+y^{2}\gg1$
and $\left|z\right|\ll1$. We can also see to what extent the intensity
of the photoelectron current is greater when the bound probability
current (green circular arrow) and the electric field (red or blue circular arrows) circulate in opposite directions. Panel
(c) of Fig. \ref{fig:j_z_summary} shows the corresponding results
for the initial state $\chi_{\rho}^{\epsilon}$. Although $\chi_{\rho}^{\epsilon}$
does not display any bound probability currents, it yields a non-zero
net photoelectron current along the $z$ direction, consistent with
its decomposition into $\chi_{\mathrm{c}}^{\epsilon}$ and $\chi_{\mathrm{c}}^{\epsilon*}$. This decomposition
along with the propensity rules for the $\chi_{\mathrm{c}}^{\epsilon}$
states dictate that the photoelectron current will flow in the direction
corresponding to the $\chi_{\mathrm{c}}^{\epsilon}$ component that
has a bound probability 
current counter-rotating with the electric field.

\begin{figure}
\noindent \begin{centering}
\includegraphics[scale=0.35]{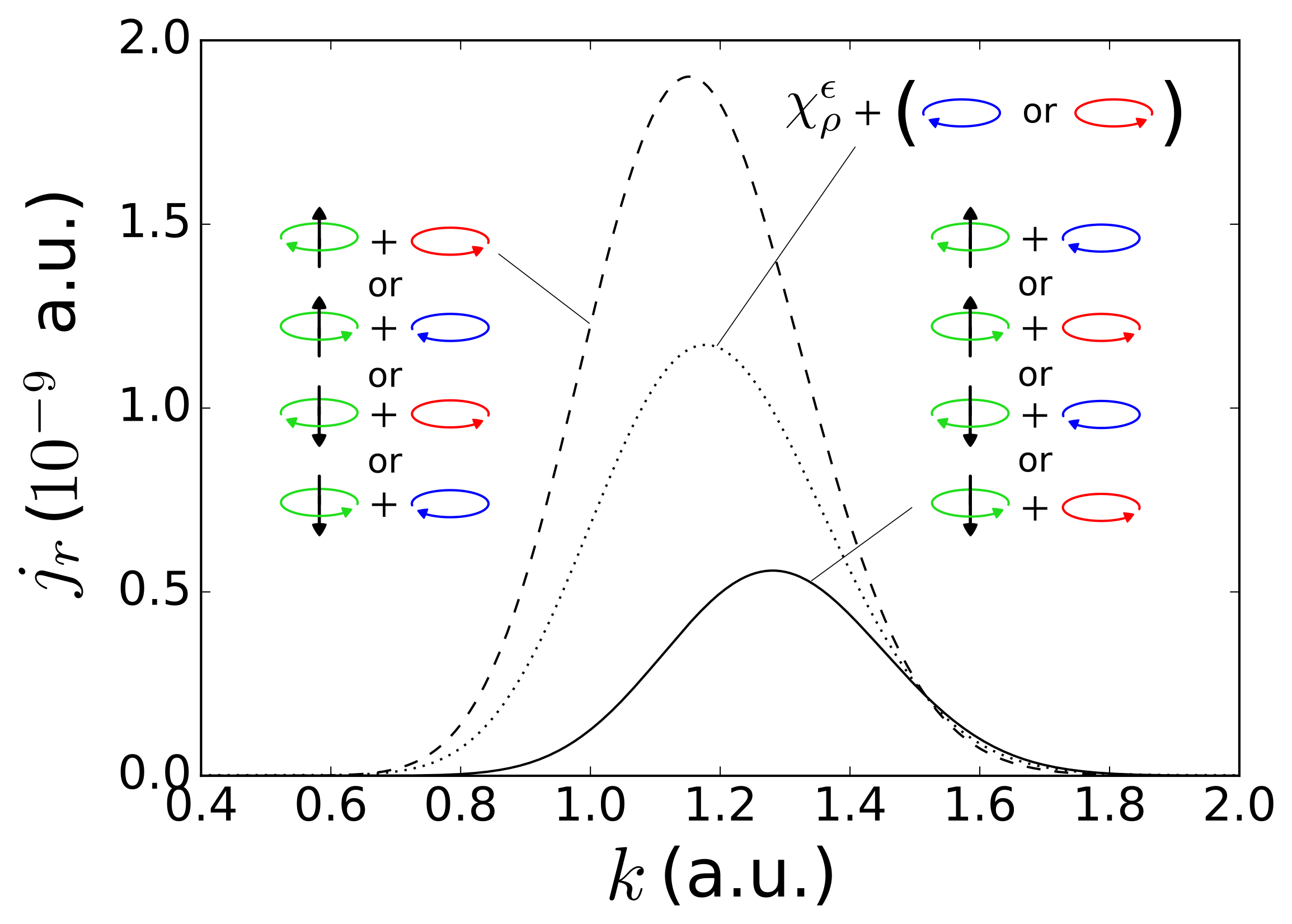}
\par\end{centering}
\caption{Radial component of the photoelectron current density (ionization
rate) as a function of photoelectron momentum, resulting from strong
field ionization of the initial states $\chi_{\mathrm{c}}^{\epsilon}$
and $\chi_{\mathrm{c}}^{\epsilon*}$ {[}see Eqs. \eqref{eq:chi_rho}-\eqref{eq:chi_c_cc}
and Fig. \ref{fig:j_z_summary} (c) for explanation of symbols{]}, via intense light circularly
polarized in the $xy$ plane. Up to $k\approx1.5\,\mathrm{a.u.}$
all the counter-rotating setups, where the azimuthal part of the bound
probability current and the light rotate in opposite directions, yield
a higher $j_{r}$. Beyond $k\approx1.5$ a.u. the co-rotating setups
yield a higher $j_{r}$. The corresponding curve for the states $\chi_{\rho}^{\epsilon}$
is simply the average between the two curves shown and is independent
of enantiomer and light polarization. The parameters $\mathcal{E}$,
$\omega$, and $I_{p}$ are the same as in Fig. \ref{fig:j_z_summary}.
\label{fig:j_r}}
\end{figure}

The marked dependence of the photoelectron yield on the relative direction
between the bound probability current and the circularly polarized
electric field (PR1) can be clearly visualized in Fig. \ref{fig:j_r},
which is the analog of Fig. \ref{fig:j_z_summary} for the radial
component of the photoelectron current. Note that for the $\chi_{\rho}^{\epsilon}$
states the total photoelectron yield $j_{r}\left(k\right)$ is independent
of both enantiomer and circular polarization.

\begin{figure}
\noindent \begin{centering}
\includegraphics[scale=0.35]{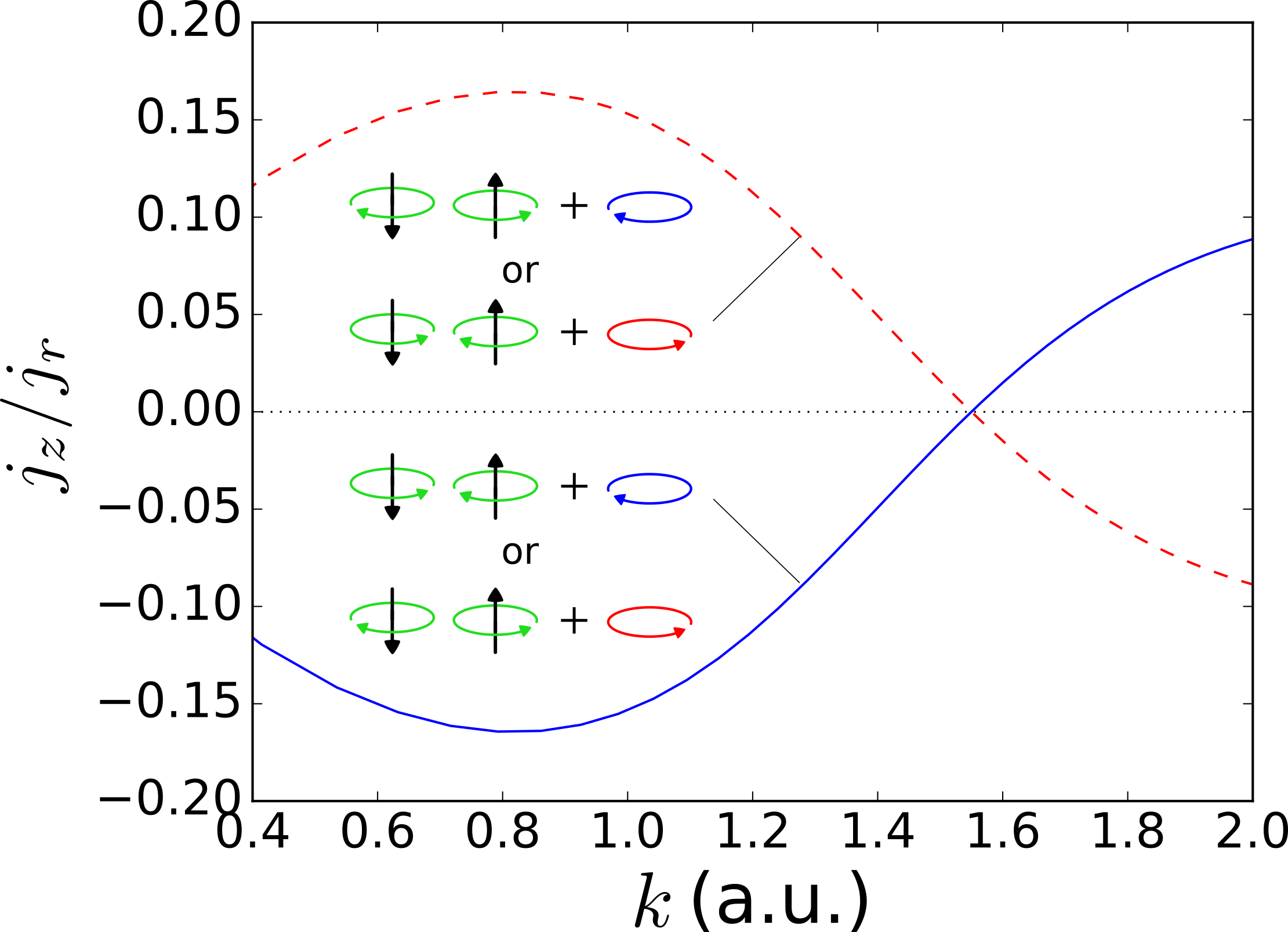}
\par\end{centering}
\caption{Ratio of the $z$ component to the radial component of the photoelectron
current density as a function of the photoelectron momentum, resulting
from strong field ionization of the initial states $\chi_{\rho}^{\epsilon}$
{[}see Eqs. \eqref{eq:chi_rho}-\eqref{eq:chi_c_cc} and Fig. \ref{fig:j_z_summary}{]},
via intense light circularly polarized in the $xy$ plane. Note that
the asymmetric part of the signal (i.e. $j_{z}$), which encodes the
dichroic and enantiomeric response, reaches up to about 15\%
of the total signal $j_{r}$. Furthermore, it changes sign for high
values of the momentum because the propensity rule for strong field
ionization of co-rotating and counter-rotating electrons is reversed there
(see Fig. \ref{fig:j_r}). \label{fig:jz/jr}}
\end{figure}

Figure \ref{fig:jz/jr} shows the ratio of the net photoelectron current
$j_{z}\left(k\right)$ to the total photoelectron current $j_{r}\left(r\right)$
released by the strong field. This ratio represents how much of the
measured signal displays enantio-sensitivity and dichroism. Its magnitude
is similar to what is typically found in the one- and few-photon absorption
case, i.e. on the order of 10\%, in agreement with recent experimental results \cite{beaulieu_universality_2016}. Figure \ref{fig:jz/jr}
also displays a clear reversal of the FBA in the high energy tail
of the photoelectron spectrum (not so evident in Figs. \ref{fig:j_z_summary}
and \ref{fig:j_r} because of the small yield at such photoelectron
momenta), which is due to the corresponding reversal of PR1 (see Fig.
3 of Ref. \cite{barth_nonadiabatic_2013}). Such a reversal was not
decidedly observed in \cite{beaulieu_universality_2016} (see Fig.
3f there), however, future experiments with access to higher repetition
rates could explore this high-momentum low-yield region of the photoelectron
spectrum to decide on the existence of this reversal. 

\section{Conclusions}
We have studied the emergence of photoelectron circular dichroism in the strong field regime by introducing a chiral initial state in the PPT formalism of strong field ionization in a simplified setting where we consider aligned molecules and ignore the effect of the anisotropic molecular potential on the photoelectrons. We derived an equation [Eq. \ref{eq:PAD_orbital_chi_rho}] for the photoelectron angular distribution that explicitly displays photoelectron circular dichroism, i.e. it contains a term which describes an asymmetry perpendicular to the polarization plane of the light and changes sign for opposite circular polarizations and for opposite enantiomers. We computed the photoelectron angular distributions for a Keldysh parameter $\gamma=0.95$ and found asymmetries of the order of ~10\%. 

We found that the mechanism and the sign of the forward-backward asymmetry in PECD in the regime of strong field ionization can be understood  as the result of the interplay of two propensity rules: (i) the strong field ionization rate depends on the relative rotation directions of the electric field and the bound electron, being higher when the electron and the electric field rotate in opposite directions \cite{barth_nonadiabatic_2011,eckart2018ultrafast,herath_strong-field_2012,Beiser_PRA_2004,Bergues_PRL_2005,Kaushal_PRA_2013,Kaushal_PRA_2013, Kaushal_2018} (ii) The `forward-backward' asymmetry depends on the direction of the current of the initial state in the region of the tunnel exit, the photoelectron is more likely to be emitted `forwards' (`backwards') if the probability current of the initial state in the tunnel exit region points `forwards' (`backwards'). For a real initial state, these two propensity rules can be applied by first decomposing the chiral probability density state into two states with opposite azimuthal currents. 

In comparison to the one-photon regime \cite{ordonez_propensity_2019,ordonez_propensity_2019_2}, we find that both propensity rules are reversed and thus the same sign of photoelectron circular dichroism is observed in the one- and in the many-photon regime for the states studied.


Excitation of chiral states in atoms \cite{nicola_mayer} and probing excited states via strong field ionization \cite{huismans_time-resolved_2011} with circularly polarized fields can also be used to verify our predictions.


\section{Acknowledgments}
We gratefully acknowledge the MEDEA
Project, which has received funding from the European
Union's Horizon 2020 Research and Innovation Programme
under the Marie Sk\l{}odowska-Curie Grant Agreement No.
641789. We gratefully acknowledge support from the DFG SPP 1840 ``Quantum Dynamics in Tailored Intense Fields'' and DFG
Grant No. SM 292/5-2.

\section{Appendix}\label{sec:Appendix}

Here we derive Eqs. \eqref{eq:PAD_orbital_chi}-\eqref{eq:D} and
\eqref{eq:PAD_orbital_chi_rho}. We begin with the derivation of Eq.
\eqref{eq:PAD_orbital_chi}. From Eqs. \eqref{eq:chi_c} and \eqref{eq:asymptotic}
we obtain the asymptotic form of $\chi_{\mathrm{c}}^{\epsilon}$,
\begin{equation}
\lim_{r\ra\infty}\chi_{\mathrm{c}}^{\pm}\left(\vec{r}\right)=\frac{\kappa^{3/2}}{\sqrt{2}}\frac{\e^{-\kappa r}}{\kappa r}\left[C_{\kappa,2}Y_{2}^{\pm1}\left(\hat{r}\right)+\i C_{\kappa,3}Y_{3}^{\pm1}\left(\hat{r}\right)\right].\label{eq:chi_c_asymptotic}
\end{equation}
Eqs. (67) and (68) in Ref. \cite{barth_nonadiabatic_2013} yield\footnote{In this appendix we will write $v_{\sigma}$ in place of $\eval{v_{\sigma}\left(t_{i}\right)}_{k=k_{n}}$
for simplicity.}
\begin{equation}
\left(v_{\sigma}^{2}+\kappa^{2}\right)\tilde{\psi}_{l,m}\left(\vec{v}_{\sigma}\right)=C_{\kappa,l}\sqrt{\frac{2\kappa}{\pi}}\left(\frac{v_{\sigma}}{\kappa}\right)^{l}Y_{l}^{m}\left(\hat{v}_{\sigma}\right)\e^{-\i l\pi/2},
\end{equation}
which we can apply to $\chi_{\mathrm{c}}^{\epsilon}$ to obtain 
\begin{equation}
\left(v_{\sigma}^{2}+\kappa^{2}\right)\tilde{\chi}_{\mathrm{c}}^{\pm}\left(\vec{v}_{\sigma}\right)=\sqrt{\frac{\kappa}{\pi}}\left\{ C_{\kappa,2}Y_{2}^{\pm1}\left(\hat{v}_{\sigma}\right)+C_{\kappa,3}\frac{v_{\sigma}}{\kappa}Y_{3}^{\pm1}\left(\hat{v}_{\sigma}\right)\right\} ,\label{eq:orb_contribution_PAD_chi_c}
\end{equation}
where we used the saddle point equation $v_{\sigma}^{2}=-\kappa^{2}$.
The formulas for the spherical harmonics $Y_{2}^{\pm1}\left(\theta,\varphi\right)$
and $Y_{3}^{\pm1}\left(\theta,\varphi\right)$ are given by 
\begin{equation}
Y_{2}^{\pm1}\left(\theta,\varphi\right)=\mp\frac{\sqrt{30}}{4\sqrt{\pi}}\sin\theta\cos\theta\e^{\pm\i\varphi},\label{eq:Y21}
\end{equation}
\begin{equation}
Y_{3}^{\pm1}\left(\theta,\varphi\right)=\pm\frac{\sqrt{21}}{8\sqrt{\pi}}\left(-5\cos^{2}\theta+1\right)\sin\theta\e^{\pm\i\varphi}.\label{eq:Y31}
\end{equation}
The polar angle of $\vec{v}_{\sigma}$ is defined through the equation\footnote{The $\pm$ in $v_{\sigma}=\pm\i\kappa$ comes from the saddle point
equation $v_{\sigma}^{2}=-\kappa^{2}$ and is unrelated to the handedness
$\epsilon=\mathrm{sgn}\left(m\right)=\pm1$ of the initial state $\chi_{\mathrm{c}}^{\epsilon}$
and to the light polarization $\sigma=\pm1$.}
\begin{equation}
\cos\theta_{v\sigma}=\frac{v_{z}}{v_{\sigma}}=\frac{k_{z}}{\pm\i\kappa},\label{eq:cos_theta_v}
\end{equation}
which in turn implies that 
\begin{equation}
\sin\theta_{v\sigma}=\sqrt{1-\cos^{2}\theta_{v\sigma}}=\sqrt{1+\frac{k_{z}^{2}}{\kappa^{2}}}.\label{eq:sin_theta_v}
\end{equation}
Using Eqs. \eqref{eq:Y21}-\eqref{eq:sin_theta_v} one can show that 
\begin{equation}
\left|Y_{2}^{\pm1}\left(\hat{v}_{\sigma}\right)\right|^{2}=\frac{15}{8\pi}\left(1+\frac{k_{z}^{2}}{\kappa^{2}}\right)\frac{k_{z}^{2}}{\kappa^{2}}\left|\e^{\pm\i\varphi_{v\sigma}}\right|^{2},\label{eq:|Y21|^2}
\end{equation}
\begin{equation}
\left|Y_{3}^{\pm1}\left(\hat{v}_{\sigma}\right)\right|^{2}=\frac{21}{8^{2}\pi}\left(25\frac{k_{z}^{4}}{\kappa^{4}}+10\frac{k_{z}^{2}}{\kappa^{2}}+1\right)\left(1+\frac{k_{z}^{2}}{\kappa^{2}}\right)\left|\e^{\pm\i\varphi_{v\sigma}}\right|^{2},\label{eq:|Y31|^2}
\end{equation}
\begin{equation}
\frac{v_{\sigma}}{\kappa}Y_{2}^{\pm1*}\left(\hat{v}_{\sigma}\right)Y_{3}^{\pm1}\left(\hat{v}_{\sigma}\right)=\frac{3\sqrt{70}}{32\pi}\left(5\frac{k_{z}^{2}}{\kappa^{2}}+1\right)\left(\frac{k_{z}^{2}}{\kappa^{2}}+1\right)\frac{k_{z}}{\kappa}\left|\e^{\pm\i\varphi_{v\sigma}}\right|^{2}.\label{eq:Y21 Y31}
\end{equation}
Equations \eqref{eq:Y21}-\eqref{eq:Y21 Y31} yield Eqs. \eqref{eq:PAD_orbital_chi}-\eqref{eq:D}.
Importantly, the FBA stems exclusively from the interference between
the two components that make up $\chi_{\mathrm{c}}^{\epsilon}$, i.e.
from the real part of Eq. (\ref{eq:Y21 Y31}). It would vanish if the
relative phase between these two components were $\pm\pi$,
which corresponds to the case where there is no probability current
along the $z$ direction in the bound state (see states $\chi_{\mathrm{p}}^{\epsilon}$
in \cite{ordonez_propensity_2019}).

Now we proceed to the derivation of Eq. \eqref{eq:PAD_orbital_chi_rho}.
The expressions for $\chi_{\rho}^{\epsilon}$ analogous to Eqs. \eqref{eq:chi_c_asymptotic},
\eqref{eq:orb_contribution_PAD_chi_c}, and \eqref{eq:PAD_orbital_chi}
read as 
\begin{equation}
\lim_{r\ra\infty}\chi_{\rho}^{\pm}\left(\vec{r}\right)=\frac{1}{\sqrt{2}}\left[\lim_{r\ra\infty}\chi_{\mathrm{c}}^{\pm}\left(\vec{r}\right)+\lim_{r\ra\infty}\chi_{\mathrm{c}}^{\pm*}\left(\vec{r}\right)\right],
\end{equation}
\begin{equation}
\left(v_{\sigma}^{2}+\kappa^{2}\right)\tilde{\chi}_{\rho}^{\pm}\left(\vec{v}_{\sigma}\right)=\frac{1}{\sqrt{2}}\left[\left(v_{\sigma}^{2}+\kappa^{2}\right)\tilde{\chi}_{\mathrm{c}}^{\pm}\left(\vec{v}_{\sigma}\right)+\left(v_{\sigma}^{2}+\kappa^{2}\right)\tilde{\chi}_{\mathrm{c}}^{\pm*}\left(\vec{v}_{\sigma}\right)\right],
\end{equation}
\begin{multline}
\left|\frac{1}{2}\left(v_{\sigma}^{2}+\kappa^{2}\right)\tilde{\chi}_{\rho}^{\pm}\left(\vec{v}_{\sigma}\right)\right|^{2}=\frac{1}{2}\bigg\{\left|\frac{1}{2}\left(v_{\sigma}^{2}+\kappa^{2}\right)\tilde{\chi}_{\mathrm{c}}^{\pm}\left(\vec{v}_{\sigma}\right)\right|^{2}+\left|\frac{1}{2}\left(v_{\sigma}^{2}+\kappa^{2}\right)\tilde{\chi}_{\mathrm{c}}^{\pm*}\left(\vec{v}_{\sigma}\right)\right|^{2}\\
+\left|\left(v_{\sigma}^{2}+\kappa^{2}\right)\right|^{2}\left[\left(\tilde{\chi}_{\mathrm{c}}^{\pm*}\left(\vec{v}_{\sigma}\right)\right)^{2}+\left(\tilde{\chi}_{\mathrm{c}}^{\pm}\left(\vec{v}_{\sigma}\right)\right)^{2}\right]\bigg\}.\label{eq:PAD_chi_rho_full}
\end{multline}
Consider how the last expression changes if we rotate the initial
wave function by an angle $\alpha$ around the $z$ axis. From Eq.
\eqref{eq:chi_c} it is evident that 
\begin{equation}
\chi_{\mathrm{c}}^{\pm}\left(\vec{r};\alpha\right)\equiv\hat{R}_{z}\left(\alpha\right)\chi_{\mathrm{c}}^{\pm}\left(\vec{r}\right)=\chi_{\mathrm{c}}^{\pm}\left(\vec{r}\right)\e^{\mp\i\alpha},
\end{equation}
i.e. the wave function acquires an overall phase factor $\e^{\mp\i\alpha}$.
Equation \eqref{eq:PAD_chi_rho_full} now takes the form 
\begin{multline}
\left|\frac{1}{2}\left(v_{\sigma}^{2}+\kappa^{2}\right)\tilde{\chi}_{\rho}^{\pm}\left(\vec{v}_{\sigma};\alpha\right)\right|^{2}=\frac{1}{2}\bigg\{\left|\frac{1}{2}\left(v_{\sigma}^{2}+\kappa^{2}\right)\tilde{\chi}_{\mathrm{c}}^{\pm}\left(\vec{v}_{\sigma}\right)\right|^{2}+\left|\frac{1}{2}\left(v_{\sigma}^{2}+\kappa^{2}\right)\tilde{\chi}_{\mathrm{c}}^{\pm*}\left(\vec{v}_{\sigma}\right)\right|^{2}\\
+\left|\left(v_{\sigma}^{2}+\kappa^{2}\right)\right|^{2}\left[\left(\tilde{\chi}_{\mathrm{c}}^{\pm*}\left(\vec{v}_{\sigma}\right)\right)^{2}\e^{\pm2\i\alpha}+\left(\tilde{\chi}_{\mathrm{c}}^{\pm}\left(\vec{v}_{\sigma}\right)\right)^{2}\e^{\mp2\i\alpha}\right]\bigg\},\label{eq:PAD_chi_rho_full-1}
\end{multline}
where only the terms on the second line depend on $\alpha$. Averaging
over $\alpha$ yields Eq. \eqref{eq:PAD_orbital_chi_rho}.

\bibliographystyle{apsrev4-1}
\bibliography{MyLibrary}

\end{document}